\documentclass[aps,pra,twocolumn,superscriptaddress,floatfix]{revtex4-2}

\usepackage{graphicx}
\usepackage{subfigure}
\usepackage{amsmath}
\usepackage{mathrsfs}
\usepackage[english]{babel}
\usepackage{amsthm}
\usepackage{amssymb}
\usepackage{hyperref}
\usepackage{xcolor}
\usepackage[OT4]{fontenc}
\usepackage[utf8]{inputenc}

\usepackage{dcolumn}
\usepackage{bm}

\hypersetup{colorlinks=true, linkcolor=black, citecolor=black, urlcolor=blue}
\allowdisplaybreaks

\begin{document}

\newcommand{\x}{{\bf r}}
\newcommand{\K}{{\bf k}}
\newcommand{\q}{{\bf q}}
\newcommand{\dk}{  \Delta {\bf k}}
\newcommand{\DK}{\Delta {\bf K}}
\newcommand{\KK}{{\bf K}}
\newcommand{\X}{{\bf R}}

\newcommand{\B}[1]{\mathbf{#1}} 
\newcommand{\f}[1]{\textrm{#1}} 

\newcommand{\half}{{\frac{1}{2}}}

\newcommand{\vv}{{\bf v}}
\newcommand{\p}{{\bf p}}

\newcommand{\dx}{\Delta {\bf r}}

\author{Pawe{\l} Zin}
\affiliation{National Centre for Nuclear Research, ul. Pasteura 7, PL-02-093 Warsaw, Poland}
\email{pawel.zin@ncbj.gov.pl}

\author{Maciej Pylak}
\affiliation{National Centre for Nuclear Research, ul. Pasteura 7, PL-02-093 Warsaw, Poland}

\title{Comment on "Absence of a consistent classical equation of motion for a mass-renormalized point charge"}

\begin{abstract}
Here we comment on the paper by Arthur D. Yaghjian, Phys. Rev. E 78, 046606 (2008) (arXiv:0805.0142). The author provides an equation of motion for a point charged particle in a certain regime of system parameters (on the other hand, claiming that in a different regime the classical equation of motion does not exist). The solutions of this equation (in the regime where it exists) presented in the paper show instantaneous jumps in the particle’s velocity. We show that such jumps, in the case of a point particle, would generate infinite energy in the radiated electromagnetic field. Therefore, we claim that the point-particle limit used by the author is incorrect. 
\end{abstract}

\maketitle

In the following we comment on the equation of motion of point charge (Eq.~(45) in the paper \cite{Yag2}).
This equation of motion is derived starting from the charged-sphere model.
Here we analyze the simplest situation considered in the paper: charge in the uniform electric field acting on it for finite time. The proposed physical situation is as follows:
"For example, the charge could be accelerated between two infinitesimally thin plates of a parallel-plate capacitor charged to produce the electric field $E_0$. It could be released at time $t = t_1 = 0$ from one plate of the capacitor and left through a small hole in the second plate at time $t = t_2$." \cite{Yag2}

When we imagine that a charged sphere enters the parallel-plate capacitor, we notice the existence of so-called "transition intervals".
These intervals occur when the external force is non-zero only on some part of the charged sphere. In the situation considered above, there are two such intervals.
The first transition interval $\Delta t_1$ starts when the external force "touches" the sphere at time $t=0$ on one side, and ends when the force "touches" the other side of the sphere - then it acts on the whole sphere. The second transition interval $\Delta t_2$ is analogous, but takes place at time $t_2$ when the external force "leaves" the sphere. 

The author of the commented paper finds that the Taylor expansion leading to the well-known Lorentz-Abraham equation does not hold during the transition intervals.
In principle, during these transition intervals, the self-force could be evaluated using the equations of motion governing the sphere (of course not using the Taylor expansion).
However, this complicated task is not undertaken by the author.
Instead, the author takes this force as unknown, having a non-zero value  only in the transition intervals. The author calls this force a transition force.
With the unknown transition forces, the equation of motion for the center velocity of the charged sphere takes the form of Eq.~(19) present in the paper \cite{Yag2}.


The author shows that the transition forces can be chosen such that the pre-acceleration and pre-deceleration are eliminated. 
These transition forces generate velocity changes across the transition intervals.
In the case of constant electric field $E_0$ acting on the charge for a certain time, these velocity changes are given by Eq.~(8.83) in \cite{Yag1}.
The velocity change across the first transition interval (which we denote by $v$) reads
\begin{equation} \label{v}
 v/c = - C_0 \ \mbox{sign}(e E_0) \log \left(  1  - \left|  \frac{eE_0 \tau_e}{mc}  \right|  \right)
\end{equation}
where $C_0 > 1$ is a constant, $e$ is the value of a particle charge, $\tau_e = \frac{e^2}{6\pi \epsilon_0 m c^3}$ and $m$ the total rest mass of the charged sphere. Here it is important to note that the velocity change $v$  {\it does not} depend on the value of $\Delta t_1$.

The final part of the paper is devoted to the point limit of the above theory which is given by the modified Lorentz - Abraham - Dirac (LAD) equation (Eq. (45) in the paper). 
The velocity changes become now instantaneous - for example for $t< 0$ the velocity is zero and for $t = 0^+$ it is equal to nonzero value $v$.

Below we focus on the solution which has this property of the instantaneous velocity change across the transition interval.
As it is known, the electromagnetic field radiated by the charge is proportional to its acceleration. When we deal with an instantaneous jump in velocity the acceleration becomes proportional to the Dirac delta function - $\delta(t)$. Thus the radiated field is proportional to $\delta(t-r/c)$.  Below we discuss the consequences of having such a field.

Now we concentrate on the problem of energy radiation in such an instantaneous pulse coming from the first transition region.
As it is known, the radiated energy is proportional to the square of the radiated part of the electromagnetic field. As a consequence, the radiated energy is proportional to the square of acceleration and reads
\begin{equation}\label{rad}
E_{rad,\Delta t} =  m \tau_e \int_0^{\Delta t} dt \, \gamma^6(t) a^2(t) > 
 m \tau_e \int_0^{\Delta t} dt \,  a^2(t).
\end{equation}
In the analyzed situation, we end up with the integral proportional to $\int dt \, \left(\delta(t) \right)^2$.

Trying to somehow define this integral, we smear the instantaneous velocity change equal to $v$ on the interval $\Delta t$.
We stress that this $\Delta t$ is NOT the transition interval for the charged sphere.
We have
\begin{equation}
\frac{1}{\Delta t} \int_0^{\Delta t} dt \, a^2(t) \geq \left( \frac{1}{\Delta t} \int_0^{\Delta t} dt \, a(t) \right)^2 = \frac{v^2}{(\Delta t)^2}
\end{equation}
where we used $ v = \int_0^{\Delta t} dt \, a(t) $.
Thus we have
\begin{equation}
E_{rad,\Delta t} >  m \tau_e \frac{v^2}{\Delta t}
\end{equation}
In the limit $\Delta t \rightarrow 0$ we obtain infinite energy radiated by the point particle.

It is crucial that the formula for the radiated energy of the charged sphere presented in the paper looks differently. In the case of energy radiated in the first transition interval it reads
\begin{equation}\label{rad2}
E_{rad,\Delta t_1} =  = \int_0^{\Delta t_1} dt \, \left(  m \tau_e \gamma^6 \dot u^2 - f_{a1} u  \right) 
\end{equation}
where $u$ is the center of sphere velocity and $f_{a1}$ is the transition force in the first transition interval.
Comparing Eqs.~(\ref{rad}) and (\ref{rad2}) we notice the presence of additional term $ - \int_0^{\Delta t_1} dt \, f_{a1} u $ in the formula for charged sphere radiated energy.
Here we shortly discuss the derivation of the above formula.

The electromagnetic energy density is a quadratic function of the electromagnetic field. In the case of point particle, the radiated electromagnetic field is a well-known function of the particle trajectory. This gives the connection between radiated energy and the particle trajectory.

In order to compute the electromagnetic energy radiated by the charged sphere in the transition interval, we would need to know the electromagnetic field radiated by the charged sphere (in the transition region). The author does not compute the radiated electromagnetic field in the transition region. 
As was written above, instead, he introduces unknown transition forces in the equation of motion of the charged sphere.
The radiated energy is computed indirectly, by simply integrating the equation of motion.

It turns out that the energy radiated in the first transition interval $E_{rad,\Delta t_1}$ is equal to (Eq.~(8.82) in \cite{Yag1})
\begin{equation} \label{Erad1}
 E_{rad,\Delta t_1} = mc^2 \left(  1 - \cosh(v/c) + \frac{eE_0 \tau_e}{mc} \sinh (v/c)    \right)
\end{equation}
where $v/c$ is given by Eq.~(\ref{v}). We present the above formulas to show that $E_{rad,\Delta t_1}$ is finite and does not depend on $\Delta t_1$. This is a crucial difference with respect to the point particle formula $E_{rad,\Delta t}$  which clearly depends on $\Delta t$ and goes to infinity as $\Delta t \rightarrow 0$.

Thus comparing point particle and charged sphere, we find that according to Yaghjian's result, the point particle emits a lot more energy than the charged sphere. Thus the charged sphere radiation, independently how small its radius is, always dramatically differs from the point charge radiation.




For a moment we forget about the infinite energy carried by such instantaneous pulse produced by the discussed velocity jump (when considering a point particle case) and consider the situation where we have two point charges.
The first charge experiences some external force for finite time and gets this instantaneous velocity jump.
This produces the above discussed field which is proportional to $\delta(t)$ - so naively speaking it has infinite value and zero width. 
This radiated field expands, finally reaching the second charge.
Clearly this radiated pulse does not satisfy inequality Eq. (46b) present in the paper \cite{Yag2}. Thus according to the paper the equation of motion of the second charge is not valid.

If we deal with the pulse that gives causal solution of the first particle  (with a jump in the velocity) it does not give a solution in the case of two particles.
It simply means that if we would try to extend the Yaghjian's theory to few particles it would not give causal theory if the first particle would have jump in velocity.
This means that the few particle causal theory cannot have this instantaneous jumps in velocity.

The above arguments show that the point charged particle theory with instantaneous velocity jumps proposed by Yaghjian is ill-defined.

\bibliography{refs}
\bibliographystyle{apsrev4-2}

\end{document}